\documentstyle[seceq,epsf]{ptptex}

\let  \q=`
\def\sqr#1#2{{\vcenter{ \vbox{\hrule height.#2pt  \hbox{\vrule width.#2pt 
height#1pt  \kern#1pt  \vrule width.#2pt}\hrule height.#2pt}}}}
 
 \def\gtsim{ \mathrel{ \hbox{\raise0.2ex
\hbox{$>$}\kern-0.75em \raise-0.9ex\hbox{$ \sim$}}}}
\def\ltsim{ \mathrel{ \hbox{\raise0.2ex
\hbox{$<$}\kern-0.75em \raise-0.9ex\hbox{$ \sim$}}}}
\def  \Romannumeral(#1) {\uppercase\expandafter{ \romannumeral#1}}

\def  \Romannumeral(#1) {\uppercase\expandafter{ \romannumeral#1}}
 \def  \Rn(#1) {\uppercase\expandafter{ \romannumeral#1}}
 \def\Fig(#1){$${\overline{\underline{ \rm Fig.{\  \ #1}}}}$$} 
\def\Tab(#1){$${\overline{\underline{
  \rm Table\  \   \uppercase\expandafter{ \romannumeral#1}}}}$$} 

 \def  \b{ \beta}

 \def  \fn1{N_{f_1}}
 \def  \fn2 {N_{f_2}}

\def  \Romannumeral(#1) {\uppercase\expandafter{ \romannumeral#1}}
\def  \Rn(#1) {\uppercase\expandafter{ \romannumeral#1}}

\title{
Two dimensional $CP^{2}$ Model with $\theta$-term and Topological Charge 
Distributions
}

\author{
Masahiro {\sc IMACHI}
\footnote{E-mail address:imachi@sci.kj.yamagata-u.ac.jp} 
\\ 
Shouhei {\sc KANOU}
\footnote{E-mail address: kanou$\underline{~}$sh@soft.hitachi.cp.jp}
and
Hiroshi {\sc YONEYAMA}
\footnote{E-mail address:yoneyama@cc.saga-u.ac.jp} 
}

\inst{
$^*$Department of Physics, Yamagata University, Yamagata 990-8560
\\
$^{**}$ Hitachi Tohoku Software Co,Sendai 980-0811
\\
$^{***}$ Department of Physics, Saga University, Saga 840-8502
}

\recdate{
}

\abst{
Topological charge distributions in 2 dimensional $CP^2$ model with 
$\theta$-term is calculated. In strong coupling 
regions, topological charge distribution is approximately given by 
Gaussian form as a 
function of  topological charge and this behavior leads to the first order 
phase transition at $\theta=\pi$. In weak coupling regions it shows 
non-Gaussian distribution and the first order phase transition disappears. 
Free energy as a function of $\theta$ shows ``flattening" behavior 
at $\theta=\theta_f<\pi$, when we calculate the free 
energy directly from  topological charge distribution. Possible origin of 
this flattening phenomena is prensented.
}

\begin{document}

\maketitle

%
%
\section{Introduction}
Quantum Chromodynamics(QCD) admits the topological term, i.e., $\theta$-term. It leads to ``strong 
CP violation''. Experimentally, the 
magnitude of  $\theta$ parameter which controls the magnitude of strong CP
 violation is severely limited to a tiny value( $| \theta|  \ltsim  10^{-9}$).
Lattice gauge theory( LGT), formulated by K.G. Wilson\cite{rf:W} and developed by 
M.Creutz\cite{rf:Cr}, has made a realistic  progress in understanding the 
nature of QCD. In LGT in which Euclidean space-time is 
adopted, usual gauge coupling is purely real quantity. On the other hand, 
$\theta$-term appears as a purely imaginary quantity. Combining both of 
these leads to complex numbered coupling. The action given by this complex 
coupling constant  gives complex valued Boltzmann weight. 
 Except for a few number of  cases, the works in LGT are done for 
 real couplings. It is, however, very important to understand the role of 
  complex coupling cases, i.e., the system with $\theta$-term. Monte Carlo
  simulations are  confronted with a difficulty due to complex Boltzmann 
   weight.    Monte Carlo simulation in this case was  made possible to some 
extent by Bhanot et al. \cite{rf:Bha}\cite{rf:BhaD}, Wiese \cite{rf:Wi} and Karliner et al.\cite{rf:K}\cite{rf:BPW}.  \par
  The method  is summarized as follows;
  1)First we obtain the topological charge \cite{rf:BL} distribution $P(Q)$ with the use of the 
  Boltzmann weight defined by the real coupling constant. Fourier series 
$$Z( \theta)=\sum_{Q} P(Q) e ^{i  \theta Q}$$
 gives the partition function $Z(\theta)$ for the system with 
  $\theta$-term. \par
  %
  2) There are some technical problems to obtain topological charge 
  distribution. Since the topological charge distribution is a rapidly 
  decreasing function, we use (i) ``set method", in which the whole range of 
$Q$ is divided into number of sets. In each set, MC simulation is performed.
  (ii)even after dividing into sets, the distribution is still rapidly 
  decreasing, so we use ``trial distribution  method", in which the Boltzmann weight 
  is normalized by some appropriate trial function so as to realize the 
  slowly changing topological charge distribution. 
  This method was applied to 2 dimensional U(1) system with 
  $\theta$-term . Wiese showed that the first order phase transition is observed at $\theta= \pi $  
   for all real couplings \cite{rf:Wi}. \par  
   Two dimensional $CP^{N-1}$ system has many qualitative similarity with
    4 dimensional QCD, like, asymptotic freedom, confinement, 
    deconfinement phase(  Higgs phase) and topological excitations( 
    instanton).  \par
    Schierholz studied 2 dimensional  $CP^3$ system with  $\theta$-term\cite{rf:SC}. 
    He observed that,  \par
    1)in strong couplings, the first order phase transition occurs at a critical 
    value $\theta_c= \pi $  and that\par
    2)in weak couplings, deviation of the position of the first order transition  from $\theta_c=\pi$  to 
    smaller values.   
      He conjectured  that $\theta_c  \rightarrow 0 $ as $\beta \rightarrow 
     \infty$( weak coupling limit) and that the continuum limit value of $\theta $ 
is zero.  \par
    If this conjecture is valid, it gives  quite important physical result, 
    but almost no confirmation  is made by other groups. One of the aim of the 
   present paper  is to investigate this issue  \cite{rf:PS}\footnote{After completion of the main part of this paper, we are informed that similar
 investigation was made by J. ~C. ~Plefka and S. ~Samuel(Phys. Rev. {\bf D56}(1997), 44). We
 would like to
  thank Burkhalter for informing about this paper.}. \par
We will briefly summarize the approach other than MC.
We analyzed 2 dimensional U(1) system with $\theta $-term \cite{rf:Col} based on the 
group 
character expansion method \cite{rf:U1}. Topological charge distribution $P(Q)$ is shown
 to be given by a Gaussian function $\exp (- \kappa_V Q^2))$ for all $\beta$. It leads to the result 
that the 
partition function is given by the  third elliptic 
theta function $\vartheta_3(\nu, \tau)$. It has a remarkable property that 
the partition function has 
an infinite numbers of zeros as a function of $z=e^{i  \theta }$, 
where $\theta$ is extended to complex values. In infinite volume limit,
 these infinite numbers of 
partition function zeros   
accumulate to $z=-1$ according to $\kappa_V \propto 1/V$.
 Namely, first order phase transition is expected at $\theta= \pi $ in the 
limit $V
 \rightarrow  \infty$, where $V=L^2$ denotes the volume of system.  \par
 
  Four dimensional $Z_N$ gauge system with $\theta$-term is discussed by 
  Cardy and Rabinovici \cite{rf:CR}. They pointed out that $\theta$-term  leads to 
  symmetries under dual transformation of Hamiltonian and this symmetry 
  leads to very rich phase structures, namely, confinement phase, Higgs 
  phase( deconfinement phase) and condensation of electric and magnetic 
  charges( oblique confinement) \cite{rf:Ho}. Their argument is based on free energy 
  and rather qualitative. It will be of much interest to investigate 
  this system by MC simulation or other method( e.g., renormalization group method). \par
    
In the previous paper \cite{rf:CP1}, we performed MC 
simulation on $CP^1$ system with $\theta$-term  in strong and weak coupling
 regions. In strong coupling 
regions, first order phase transition is observed at $\theta= \pi$ \cite{rf:FB}\cite{rf:IPZ}.
 In weak coupling regions \cite{rf:Sei}, we observed that  topological charge distribution  
 deviates apparently from Gaussian 
 form. Free energy in weak coupling regions is 
 approximately given by $\cos  \theta$ form, which shows only a few small 
 $Q$'s control the 
 topological charge distribution. In intermediate coupling regions, obtained results
  were not  so clear.  In order to avoid the effect of statistical error as
 far as possible, we used fitted analytic form for $P(Q)$ to obtain 
$Z(\theta), F(\theta)$ etc. Then even in weak coupling region we observed 
no flattening behavior but change of behavior of $Z(\theta), F(\theta)$. 
The power series fit to $P(Q)$ has the property; \par
 1)for $\b$  small( strong coupling regions), it is simply given by 
 Gaussian form( $\exp (- \kappa_V Q^2))$.  \par
 2)for $\b$  large( weak coupling regions), we need $Q^4$ term in the 
 polynomial fit of the exponent of $P(Q)$.  \par
 3)for $\b$  intermediate, power series fit suffers from quite large 
 $\chi ^2$ value.  \par

 
 \par
 In this paper, CP$^2$ model with $\theta$-term will be studied by MC method \cite{rf:IKnY}.
  The emphasis is put on intermediate coupling 
 regions($  \b=3  \sim 4$). From $P(Q)$ we obtain $Z(\theta)$ and 
  we obtain free  energy per unit volume by $F(\theta)=-(1/V) \ln Z( \theta)$.
 $F(\theta)$ grows smoothly in smaller $\theta$ and shows ``flattening'' 
 at some value $\theta_f$, beyond which $F(\theta)$ becomes almost flat. For 
 example, at $\b=3.5$ and $ \ L=20$( 100000=100 kiro iterations), 
flattening is 
 observed at $\theta= \theta_f \sim 0.57\pi$. Does it correspond to the ``first
  order phase transition'' found by Schierholz in CP$^3$ system? \par

 Does the flattening mentioned above really mean a phase transition?
 We should be careful about the statistical error induced by the 
 simulation.
For example, after $N$ times of sweeps( =iterations), we would have statistical error 
of order 
   $$ \delta P = |{\rm error}|  \sim 1/  \sqrt N  .$$
Note that $\delta P$ is defined as positive.

\begin{figure}
	\epsfxsize= 4 cm
	\centerline{\epsfbox{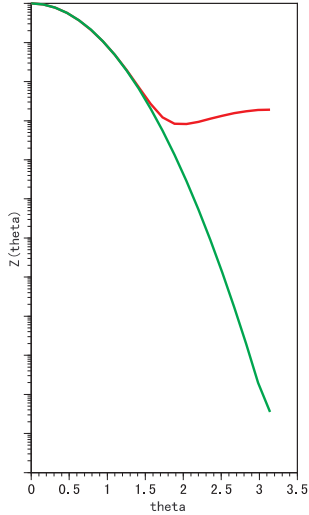}}
\caption{An illustration of partition function as a function of theta. 
Direct fourier series( curve with flattening behavior) and one( Gaussian curve) calculated from smooth fit to $P(Q)$.       }
   \label{z-b10}
\end{figure}

Set method  and trial function method will give us $P(Q)$ containing 
great number of orders of decrease from $Q=0$ to large value of $Q$. 
If observed $P(Q)$ is fitted by a smooth function
  $P_S(Q)$, it can be expressed as,
  $$P(Q)=P_S(Q)+ \Delta P(Q),$$
  where $\Delta P(Q)$ is the difference between observed $P(Q)$ and the fitted 
  function $P_S(Q)$. Quantity $| \Delta P(Q)|$ will be approximately given by 
  $  \delta  P  \times P(Q)$.
   The quantity $\Delta P(Q)$ can be both positive and 
  negative( or zero), since it is a fluctuation. 
 The value 
of $  |\Delta P(Q)|$ is approximately of order of $(1/ \sqrt N) P(Q)$ and 
$  \Delta P(Q)$ itself 
is expected to be small at large $Q$. Partition function $Z(\theta)$, 
 however, is given by 
the fourier series and the error in evaluation of $Z(\theta)$ is controlled by the largest one, i.e.,
 $|\Delta P(0)|  \sim   \delta P  \sim 1/ \sqrt N$.  
  Topological charge distribution $P(Q)$ is a rapidly decreasing function.
 When we estimate $Z(\theta)$ from  
$$Z( \theta)=\sum_{Q} P(Q) e ^{i  \theta Q},$$
 the value of $Z(\theta)$  will  also be a rapidly decreasing function of $\theta$ and 
  the quantity $\Delta P(0)$ will play an important role at large $\theta$.
  Since the partition function obtained by $P_S(Q)$ is  a rapidly decreasing 
  function,   true $Z_S(\theta)$ will be masked by 
  the error  $\Delta P(0)$ at large $\theta$,
   which is much larger than $P_S(Q)$ itself 
  at large $Q$ . 
  This effect leads to the shape of $Z(\theta)$ with flattening at some 
  $\theta_f$ just like the illustration shown in Fig.\ref{z-b10}($\theta_f\sim 0.55\pi$ in this figure).

The value of  $Z(\theta)$ will almost coincide with $Z_S(\theta)$  
in $  \theta  <  \theta_f $ 
but 
 it deviates far from $Z(\theta)_S$ in $  \theta \gtsim \theta_f $. 
$Z_S(\theta)$ will become quite small  as $\theta  \rightarrow  \pi$ 
but obtained $Z(\theta)$
  does not show such suppression.  \par
  
  In this way the combined method of set and trial function leads to 
satisfactory result for $P(Q)$ but we still encounter  difficulty in 
obtaining correct $Z(\theta)$ at large $\theta$'s. \par
  In order to confirm the physical flattening (if it exists), we need
enough 
  accuracy of $Z(\theta)$ over huge range of magnitude.  
  For example to confirm the behavior of partition function down to 
10$^{-8}$, 
  we need  $N  \sim 10^{16}$ iterations.  \par
  To obtain information about flattening within available computer 
facilities,
 we should analyze at larger $\beta$ or smaller volume $V$,
since $P(Q)$ decreases faster at $\beta=$ large or at $V=$ small, partition 
function $Z(\theta)$, which is a Fourier transform of  $P(Q)$, falls slowly
 at $\beta=$ large or $V=$ small. In these regions $Z(\theta)$ will be  
 reliably estimated over whole 
 region of $\theta (0 \le \theta \le \pi$).

\section{ $ CP^2 $ model in two dimensions}
\subsection{ $ CP^2 $ model in two dimensions}
We define the $CP^2$ model with $\theta$-term
in the two dimensional euclidean lattice space. 
The $CP^2$ model consists of  three component complex scalar fields, 
\begin{eqnarray}
z_{\alpha}(n) = 
  \left(
     \begin{array}{c}
        z_1 (n) \\
        z_2 (n)  \\
        z_3 (n)  \\
       \end{array}
  \right),   
\end{eqnarray}
at each site $n$, and constrained by
\begin{eqnarray}
\sum_{\alpha = 1}^{3} z_\alpha^\ast(n) z_\alpha(n) = 1 . 
\end{eqnarray}
Action,  which is  U(1) gauge invariant,  is given by
\begin{eqnarray}
S_{\theta}[z] &=& S[z] - i \theta Q[z]  \\
S &=& \beta \sum_{n=1}^{L^2} \sum_{\mu = 1, 2}
\{ 1 - \sum_{\alpha=1}^3 {|z_\alpha^{\ast}(n) z_\alpha(n+\mu)|}^2 \},  
\end{eqnarray}
where $Q$ is topological charge,  and $z_\alpha(n)$ is coupled with 
$z_\alpha(n+\mu)$
at the nearest neighbor site $ n+\mu(\mu =1, 2) $. 
Topological charge $Q$ is defined by
\begin{eqnarray}
Q &=& \frac{1}{4\pi} \sum_{plaq} \epsilon_{\mu\nu} 
( A_{\mu}(n) +A_{\nu}(n+\mu) -A_{\mu}(n+\nu)-A_{\nu}(n) ) 
\end{eqnarray}
where,  $A_{\mu}(n)= \arg (z^{\ast}(n) z(n+\mu))$. 
Topological charge $Q$ is given by the winding number, i.e., how often the field 
$A_{\mu}(n)$ covers $U(1)$ space when we move around the periodic boundary of the whole
 space time once. 

The partition function as a function of the coupling constant $\beta$
and $\theta$ is
\begin{eqnarray}
Z( \theta ) = 
  \frac{\int D z^{\ast} D z \exp(-S[z ] + i\theta Q[z])}{\int D z^{\ast} D 
z \exp(-S[z])}
\end{eqnarray}  
where $D z $ denotes integration over  all fields, 
and the free energy $ F(\theta)$ is given by 
\begin{eqnarray}
F(\theta) = -\frac{1}{V} \ln Z(\theta), 
\end{eqnarray}
where $V$ is volume of the lattice space. 

We would like to compute $\theta$ vacua effects, 
but the complex Boltzmann factor $ e^{i\theta Q}$
prevents us from performing  ordinary Monte Carlo simulation.  In order to 
get over 
this difficulty, we follow Wiese's idea. The updating is
restricted to the fields in the topological sector. Thus the phase factor
$ e^{i\theta Q[z]}$ is replaced by a constant and can be factored out from 
the functional  integral. We can perform ordinary Monte Carlo 
simulation using the real action $S[z]$. The partition function is given 
by the 
summation of the topological
charge distribution $ P(Q) $ weighted by $ e^{i\theta Q}$ in each $Q$ 
sector. 
In practice, we get the topological charge distribution $ P(Q) $ at 
$\theta =0$
by the Monte Carlo simulation. We obtain the partition function $ 
Z(\theta) $ by taking the 
Fourier transform of $P(Q)$. 
\begin{eqnarray}
Z( \theta )   &=& \sum_{Q} e^{i \theta Q } P(Q)  
\label{eq:pq}
\end{eqnarray}
where P(Q) is
\begin{eqnarray}
  P(Q) &\equiv& 
  \frac{\int D z^{\ast(Q)} D z^{(Q)} \exp(-S[z ] )}{\int D z^{\ast} D z 
\exp(-S[z])}
\end{eqnarray}  
The integration measure $ D z^{(Q)} $ is restricted to the fields in the 
topological sector 
labeled by the topological charge $Q$. 
Note that $ \sum_{Q} P(Q) = 1 $. 

The expectation value of an observable $ O $ is given in terms of $P(Q)$ as
\begin{eqnarray}
\langle O \rangle_{\theta} &=&
 \frac{\sum_{Q} P(Q) \ \langle O \rangle_{Q} \ e^{-i \theta Q}}{\sum_{Q} 
P(Q)\  e^{-i \theta Q}},  
\end{eqnarray}
where $ \langle O \rangle_{Q} $ is the expectation value of $O$ at 
$\theta=0$
 for a given $Q$ sector 
\begin{eqnarray}
\langle O \rangle_{Q} =
 \frac{\int D z^{\ast(Q)} D z^{(Q)} \ O \ e^{-S[z]}}{\int D z^{\ast(Q)} D 
z^{(Q)} \ e^{-S[z]}}.  
\end{eqnarray} 
%

\subsection{Algorithm}
We measure the topological charge distribution $P(Q)$ by Monte Carlo
simulation with the Boltzmann weight exp($-S$), where $S$ is defined by 
eq.(2.4). 
The standard Metropolis method is used to update configurations. 
To calculate $P(Q)$, we count the number of times the configuration of $Q$
is visited by histogram method.  The distribution $P(Q)$  damps 
very rapidly as $ \mid Q \mid $ becomes large. We need to calculate $P(Q)$
at as larger $|Q|$ as possible, 
which would contribute to $ F(\theta), \langle Q \rangle_{\theta}$ and
$\langle Q^2 \rangle_{\theta}$ because they are obtained by 
Fourier transformation of $P(Q)$ and the  derivatives of the partition 
function by $\theta $. 
 In order to obtain  $P(Q)$ at larger $Q$'s, we apply following  two 
techniques, 
\begin{enumerate}
   \item[a)] the set method,
   \item[b)] the trial distribution method.
\end{enumerate}   
\vskip .5cm
{\sl The set method };\\
Range of whole $Q$ is divided into number of sets $S_i \ (\ \ i=1, 2, 
\cdots )$. Monte Carlo updatings are done in each  set $S_i$
$(S_i =\{ Q\mid 3(n-1) \le Q \le 3n \} )$. In the process, we start from a 
configuration within the set $S_i$, and we produce a tentative 
configuration $C_t$. 
When Q of this  tentative configuration $C_t$ stays within one of the bins 
in $S_i$,  the configuration $C_t$ is accepted, and the count of the
corresponding $Q$ value is increased by one. On the other hands, 
when $C_t$ goes out of the set $S_i$, $C_t$ is rejected, and
 the count of the $Q$ value of the old configuration is increased
 by one. This is done for all sets $S_i(\ i=1, 2, \cdots)$. \\
 
{\sl The trial distribution method};\\
Topological charge  distribution $P(Q)$ is still  rather rapidly 
decreasing function of $Q$ even in each set, the number of count in each 
set is sharply peaked at smallest $Q$ and almost no counts in the larger 
$Q$'s, so  statistical weight  is  modified by introducing trial 
distributions $P_t(Q)$ for each set. Namely Boltzmann weight is replaced 
by $\exp[ - S]\ / P_t(Q)$.
 This is to remedy $P(Q)$ which falls too rapidly even within
 a sets in some cases. We make the counts at $Q=3(i-1), \ 3(i-1)+1, 
 \ 3(i-1)+2$ and $3i$ in each set $S_i$ become almost the same. As the 
 trial distributions $P_t(Q)$, we apply the form
\begin{eqnarray}
P_t(Q) = A_i \exp[-{\frac{C_i(\beta)}{V} }Q^2]
\end{eqnarray} 
where the value of $C_i(\beta)$ and
$A_i$ depend on the set $S_i$. That is, the action during updating is 
modified 
to an effective one such as $S_{\rm eff} = S + \ln P_t(Q)$. 

To reproduce a normalized distribution $P(Q)$ in the whole range of $Q$
 from the counts at each set, we make matching as follows:
\begin{enumerate}
\item[i)] At each set $S_i(i=1, 2\cdots)$, the number of counts is
multiplied by $P_t(Q)$ at each $Q$. We call the multiplied value
$N_i(Q)$, which is hopefully proportional to the desired topological
charge distribution $P(Q)$. 
\item[ii)] In order to match the values in two neighboring sets $S_i$
 and $S_{i+1}$,  we rescale $N_{i+1}(Q)$ so that $N_{i+1}(Q) \to 
 N_{i+1}(Q) \times r $, where $r= N_i(Q=3i) /N_{i+1}(Q=3i)$, 
 the ratio of the number of counts at the right edge of $S_i$ to
  that at the left edge of $S_{i+1}$. These manipulations are performed 
  over all the sets. 
\item[iii)] The rescaled $S_i$'s are normalized to obtain $P(Q)$
such that
\begin{eqnarray}
P(Q) = \frac{N_i(Q)}{\displaystyle{\sum_{i} \sum_{Q} N_i(Q)}}.
\end{eqnarray}
\end{enumerate}

\section{ Numerical result I}
We use square lattice with periodic boundary condition. We measure
$P(Q)$ in various lattice sizes $(V=L^2)$ and coupling constants $(\beta)$. 
The error analysis is discussed in the previous paper\cite{rf:CP1}. 
To check the algorithm,  we calculated the internal energy. It agrees with
the analytical results of the strong and weak coupling expansions. 
Using the calculated $P(Q)$, we will estimate the free energy $F(\theta)$
 and its derivative $ \langle Q \rangle_{\theta} $, respectively. 
To obtain $Z(\theta)$, we made two methods.
\begin{itemize}
\item[(i)]We use  the measured $P(Q)$ directly to obtain $Z(\theta)$ (``direct method").
\item[(ii)] To avoid the error problems discussed in the introduction,  We first fit the measured $P(Q)$ by the appropriate function
$P_{\rm S}(Q)$, and then obtain $Z(\theta)$ by Fourier transforming from 
$P_{\rm S}(Q)$(``fitting method"). 
\end{itemize}
%
\subsection{ Topological charge distribution $P(Q)$}
\begin{figure}
\epsfxsize= 10 cm
\centerline{\epsfbox{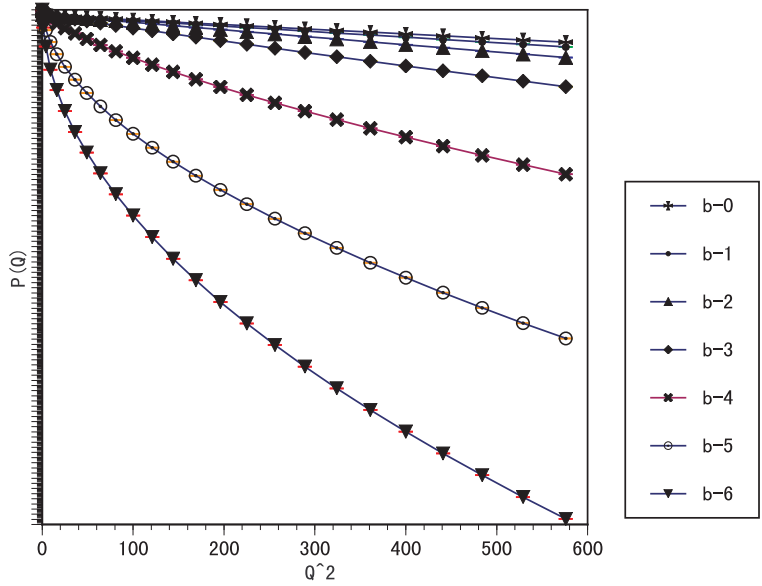}}
\caption{Topological charge distribution $P(Q) {\rm vs.}  Q^2$. 
The lattice size is $L=20$. $ \beta$ is from $0$ to $6$. 
The results include errors and are plotted for $ 0 \le Q \le 24 $. 
The total number of counts $N$ in each set is $10^5$. $P(Q)$ ranges 1 to $10^{-98}$ for $\beta=6$ case.}
\label{pq}
\end{figure}
\begin{figure}
	\epsfxsize= 10 cm
	\centerline{\epsfbox{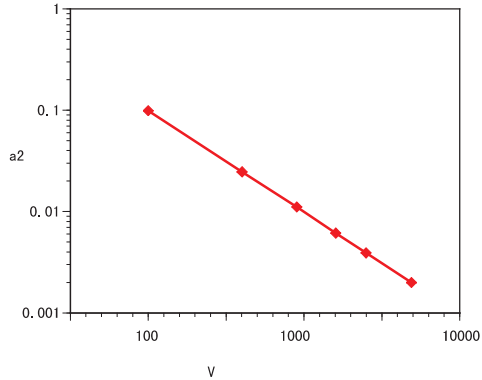}}	
   \caption{$ \ln a_2 $ vs. $\ln V$ for $ \beta = 1.0 $ and $N=10^5$. }
\label{b1-v-law}
\end{figure}
\begin{figure}
	\epsfxsize= 10 cm
	\centerline{\epsfbox{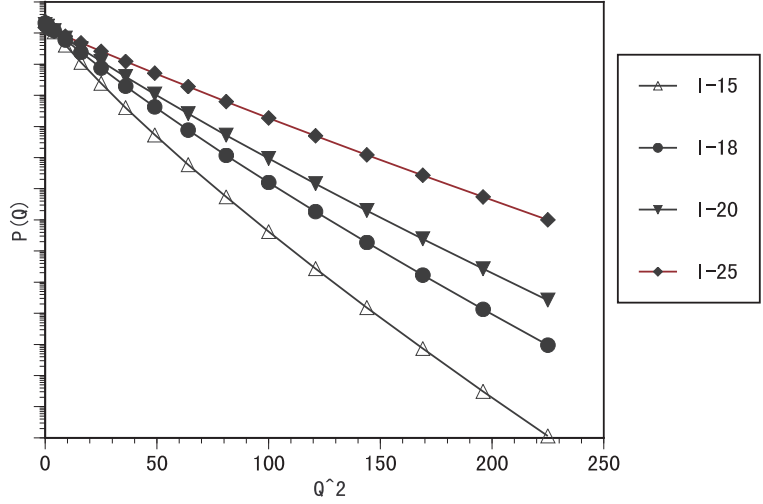}}	
   \caption{$P(Q)$ vs. $Q^2 $ for $\beta=3.45$ and  $N=5*10^5$.}
    \label{pq345}
%
\end{figure}

In this subsection we discuss the topological charge distribution $P(Q)$. 
In Fig. \ref{pq}, we show the measured $P(Q)$ for various $\beta$'s for a 
fixed volume $(L=20)$. These $P(Q)$'s have a different behavior between 
the strong coupling regions and the weak coupling regions. 
In strong coupling regions, $P(Q)$ exhibits Gaussian behavior. 
In weak coupling regions, $P(Q)$ deviates gradually from the Gaussian form. 
In order to investigate the strong coupling regions in detail,  
we use the chi-square-fitting to $\ln P(Q)$. Table \ref{table:chi2} 
and \ref{table:chi4}  display
the results of the fittings, i.e., the coefficients $a_n$ of the polynomial
$\sum_n a_n Q^n $, for various $\beta$'s with the resulting $\chi^2/$ dof,
 where dof means 
the degree of freedom. 
(i)For $\beta \le 1. 5$, the $P(Q)$'s are fitted well by the Gaussian form. 
(ii)For $ 2 \le \beta \le 3 $, terms up to the quartic one are needed for
sufficiently good fitting. 
(iii)For $ 4 \le \beta $, the fittings using the quartic polynomial
turn out to be quite poor. 
Here we discuss the volume dependence. In strong coupling regions, 
$P(Q)$ is fitted very well by  Gaussian form for all values of $ V$,
\begin{eqnarray}
P(Q) \propto \exp (- \kappa_V(\beta) Q^2),  
\end{eqnarray}
where the coefficient $ \kappa_V(\beta) (=a_2)$ depends on $\beta$ and
$V$. Fig.\ref{b1-v-law} shows $ \ln\kappa_V(\beta) $ vs. $\ln V$ for
a fixed $\beta(=1. 0)$. We see that $\kappa_V(\beta)$ is clearly proportional to
1/$V$
\begin{eqnarray}
\kappa =  \frac{\alpha}{V} 
\end{eqnarray}
This 1/$V$-dependence of the Gaussian behavior determines the phase structure
of the strong coupling regions. This result agrees with the $CP^1$ result. 
%
Fig.\ref{pq345} shows the volume dependence of $P(Q)$ for $L=15, 18, 20, 25 $
 at $ \beta=3.45$ .  Parameters $a_1^{\gamma}, \gamma=2.0 $ at $\beta=3.45$ (see Table.\ref{table:gamma}) give $V$-dependence. We  find slight deviation from exact  1/$V$ law, 
  but a clear volume dependence is observed, which gives  $\sim 1/V^{1.15}$, not so far 
 from 1/$V$ law. 
\begin{table}[h]
\caption[chi]{The results of chi-square-fitting to $\ln P(Q)$ in terms of
the polynomial $a_0 + a_2 Q^2 $ for various $\beta$,
where dof means degree of freedom.}
\label{table:chi2}
\begin{center}
\begin{tabular}{c|c|c|c|c}
\hline
\hline
$\beta$ & $\chi^2$ & $\chi^2/$dof &$ a_0 $&$ a_2 $ \\
\hline 

 $0$  & $20. 058$ & $0. 872$ & $-2. 509$ & $-0. 021$  \\
 $1$  & $83. 936$ & $3. 649$ & $-2. 416$  & $-0. 025$ \\
 $ 1. 5 $ & $136. 066 $&$ 5. 916 $& $-2. 368$ &$ -0. 028 $ \\
 $2$  & $769. 635$ & $34. 625$ & $-2. 296$ & $-0. 033$ \\
 \hline
\end{tabular} 
\end{center}
\end{table}
\begin{table}[h]
\caption[chi]{The result of chi-square-fitting to $\ln P(Q)$ in term of
 the polynomial $ \sum_n a_n Q^n $ for various $\beta$. }
\label{table:chi4}
\begin{center}
\begin{tabular}{c|c|c||c|c|c|c|c}
\hline
\hline
$\beta$ & $\chi^2$ & $\chi^2/$dof &$ a_0 $&$ a_1$&$ a_2 $& $a_3$ &$ a_4$ \\
\hline
 $2$  & $27. 835$ & $1. 392$ & $-2. 273$ & $0. 006$ & $-0. 035$ & $5. 123*10^{-5}$ & 
 $1. 359*10^{-7}$ \\
 $3$  & $30. 561$ & $1. 528$ & $-1. 950$ & $0. 01$ & $-0. 069$ & $5. 765*10^{-4}$ &
                                                     $-2. 904*10^{-6}$ \\
 $4$  & $3496. 599$ & $174. 830$ & $-0. 791$ & $-0. 684$ & $-0. 188$ & 
 $5. 842*10^{-3}$ & $-8. 643*10^{-5} $ \\
 $5$  & $4027. 247$ & $201. 362$ & $4. 147*10^{-3}$ & $-4. 468$ & $-0. 159$ &
  $6. 161*10^{-3} $ & $-1. 015*10^{-4} $ \\
 $6$  & $5175. 739$ & $258. 787$ & $-3. 341*10^{-3}$ & $-8. 806$ & $-0. 054$ & 
 $1. 423*10^{-3} $ & $-1. 393*10^{-5} $  \\
\hline
\end{tabular} 
\end{center}
\end{table}
%
\subsection{Free energy and expectation value of topological charge}
\begin{figure}
	\epsfxsize= 10 cm
	\centerline{\epsfbox{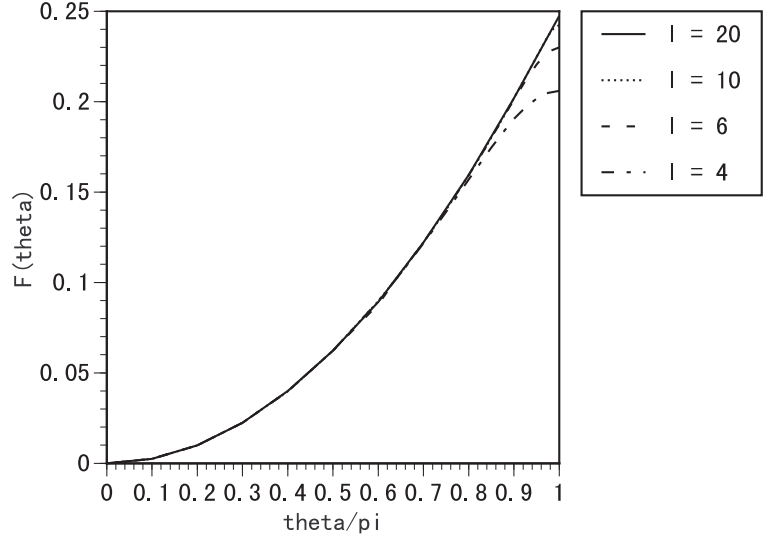}}
	\caption{Free energy vs theta in strong coupling regions. $L=4, 6, 10, 20$   }
    \label{f-b1}
\end{figure}

\begin{figure}
	\epsfxsize= 10 cm
	\centerline{\epsfbox{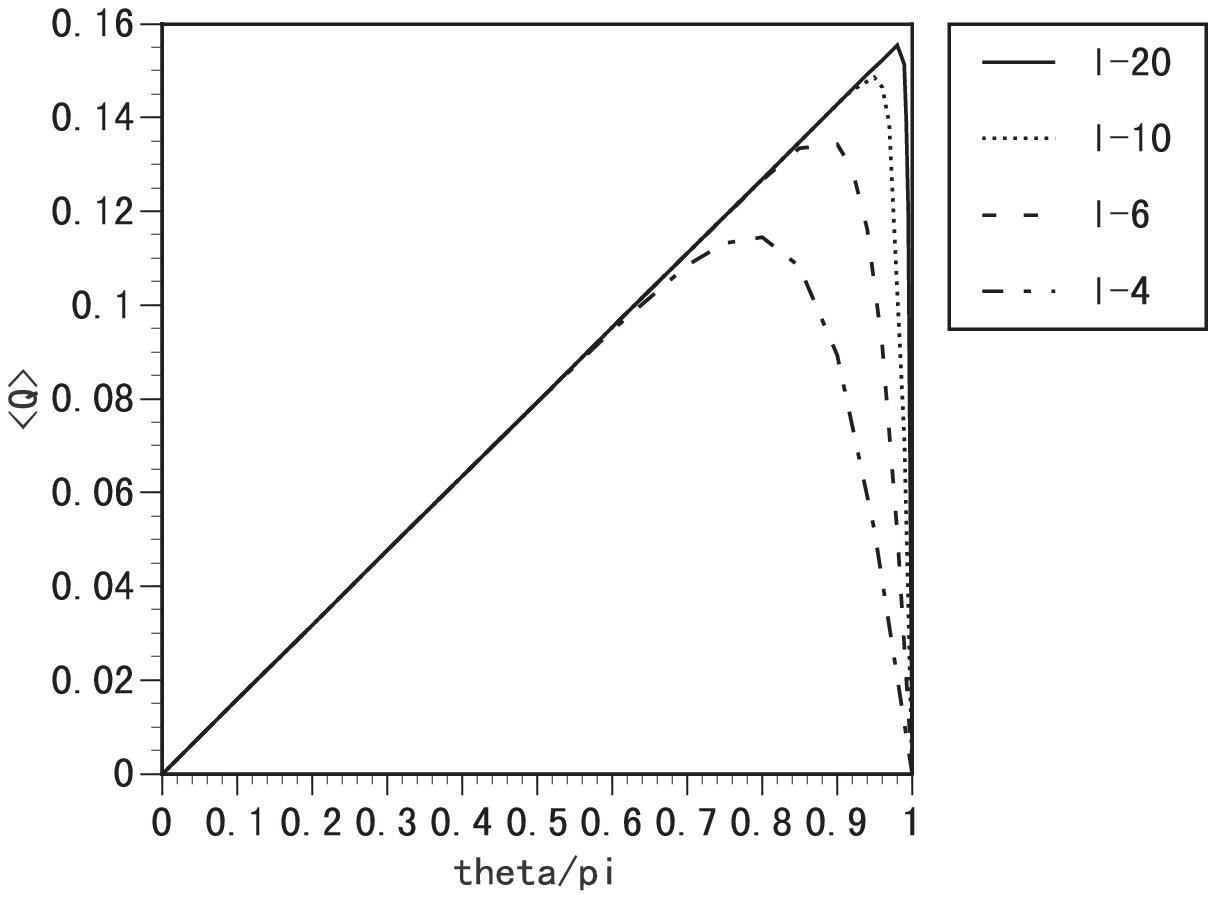}}
	\caption{$<Q>$  vs theta in strong coupling regions. $L=4, 6, 10, 20$     }
    \label{fd-b1}
\end{figure}

  The partition function $ Z(\theta)$ is given as a function of $\theta$
by (\ref{eq:pq}) from $P(Q)$. The free energy is
\begin{eqnarray}
F(\theta) &=& - \frac{1}{V} \ln Z(\theta) .         
\label{eq:F}
\end{eqnarray}
The expectation value of topological charge is
\begin{eqnarray}
\langle Q \rangle_{\theta} &=& -(-i)\frac{\displaystyle{dF(\theta})}{\displaystyle{d \theta}}       
\label{eq:Q}
\end{eqnarray}
In strong coupling regions, we have  Gaussian behavior of $P(Q)$, and
the 1/$V$-law appears to hold up to $L=70$. It is natural to expect that
this behavior persists to $V \to \infty $. Let us look at how the 1/$V$-law
affects $F(\theta)$ and $\langle Q \rangle_{\theta}$. 
By putting $ C(\beta)= 9. 9  $, in $P(Q) = \exp[-\frac{C(\beta)}{V} Q^2] $
for $\beta = 1. 0$, we calculate $F(\theta)$ and 
$ \langle Q \rangle_{\theta} $ from (\ref{eq:pq}), (\ref{eq:F}) and (\ref{eq:Q}).
 Figure.\ref{f-b1} and \ref{fd-b1} show their volume dependence. 
%
 %
As $V$ is increased, $F(\theta)$ very rapidly approaches a quadratic 
from in $\theta$ from below. 
Its first moment $\langle Q \rangle_{\theta} $ develops a peak near 
$ \theta = \pi$, and the position of the peak quickly approaches $\pi$
as $V$ increases. The jump in $\langle Q \rangle_{\theta} $ would arise at 
$\theta = \pi$ as $V \to \infty $. This indicates the existence of the
first order phase transition at $\theta = \pi $. 

\section{ Numerical results   II}
%
\subsection{ Flattening of free energy}
\begin{figure}
	\epsfxsize= 10 cm
	\centerline{\epsfbox{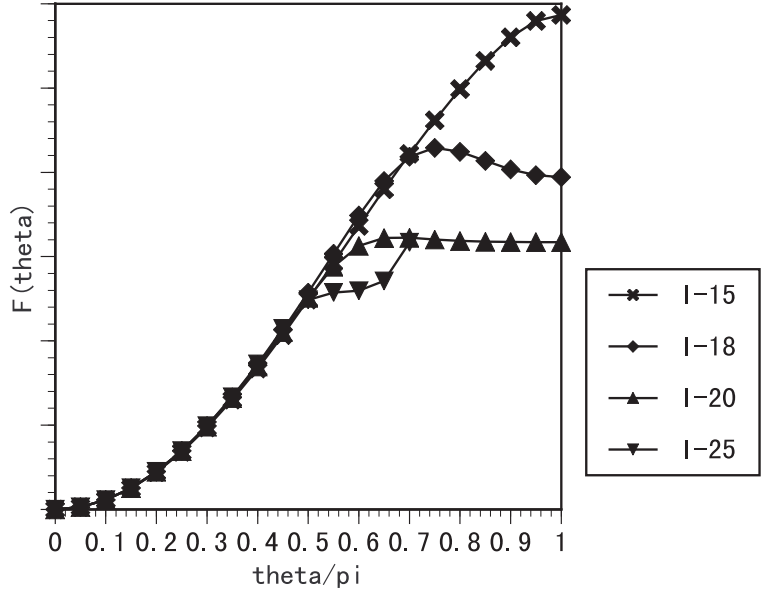}}
	\caption{Free energy vs theta at $\beta=3.45$ and  $L=15, 18, 20, 25$. Flattening is clearly 
	seen in $L=18, 20, 25$.   }
    \label{f-b345-b}
\end{figure}

 Figure \ref{f-b345-b} shows $F(\theta)$ obtained by ``direct method" at 
$\beta=3. 45$. 
$F(\theta)$ shows ``flattening '' at some value $\theta_f$ over $L=18$. 
For $\theta \le \theta_f$, $F(\theta)$ is volume independent. 
For $ \theta_f \le \theta $, $F(\theta)$ is flat. $\theta_f$ decreases 
as $V \to $ large. 
(In $L=25$, we can not calculate $F(\theta)$ for $\theta \ge 0.7\pi$, 
because $Z(\theta)$ becomes negative.) 
This flattening phenomena may correspond to similar behavior found by 
Schierholz.
According to him, such flattening is the position of the first order phase 
transition
( because $dF/ d\theta$ is ``discontinuous" at the sharp flattening of 
free energy  $F(\theta)$. Our interpretation about the flattening found in 
our numerical calculations is different.   The shift of $\theta_f$ ( 
flattening position ) with the size of the volume in our calculations is 
closely related to the statistical error existing in the estimate of 
$Z(\theta)$ through fourier series. The magnitude of statistical error 
$\delta P$ in the simulation of $N$ measurements is essentially given by 
\begin{equation}
      \delta P\sim \frac{1 }{\sqrt{ N }}.
\end {equation}
Statistical error $\Delta P(Q)$ at $Q$ in topological charge distribution 
$P(Q)$ decreases rapidly with $Q \rightarrow $ large in the set method,
\begin{equation}
      \Delta P(Q) \sim \delta P \times P(Q)
\end {equation}
where $\delta P=$ constant, and then $\Delta P(Q)$ is a rapidly decreasing 
function of $Q$.
The largest contribution of $\Delta P(Q)$ comes from that at $Q=0$,
\begin{equation}
      \Delta P(0) \sim \delta P \times P(0).
\end {equation}
Then $Z(\theta)$ suffers from the error of $P(0)$ with the magnitude of 
$\Delta P(0) \sim \delta P$. The partition function $Z(\theta)$ is also a 
rapidly decreasing  function of $\theta$ and  then $Z(\theta)$ becomes  
the order of $\Delta P(0) \sim \delta P$ at $\theta \sim \theta_f$. In 
such case, rapidly decreasing function $Z(\theta)$ at $\theta > \theta_f$ 
can not be estimated safely. \par
We will present the results of our simulation. 
In table \ref{table:theta-f} we show the flattening position 
$\theta_f$ for various lattice sizes $L$ and coupling constants $\beta, 
\quad \beta=3.45 $ and $ \beta=3.5$.

\begin{table}[h]
\caption[thf]{The  flattening position $\theta_f$ for various lattice 
sizes $L$. We also show the values of $\theta_f \times L$.}
\label{table:theta-f}
\begin{center}
\begin{tabular}{l|c|r|r}
\hline
\hline
$\beta$  & $L$  & $\theta_f$    & $ \theta_f \times L $ \\
\hline 
 3.45    & 15   &    0.83$\pi$  & $12.4\pi$             \\
         & 18   &   $0.7\pi$    & $12.6\pi$   \\
         & 20   &  $ 0.6\pi $   & $12.1\pi$     \\
         & 25   &  $ 0.51\pi $   & $12.8\pi$     \\
 3.5        & 20   &  $ 0.57\pi $   & $11.4\pi$     \\
            & 30   &  $ 0.38\pi $   & $11.4\pi$     \\
 \hline
\end{tabular} 
\end{center}
\end{table}
From these results, we observe that the flattening position 
scales as 
$\theta_f \propto L^{-1}=V^{-1/2}$ for each $\beta$, where $V=L^2$. What is the reason for 
this $V^{-1/2}$ -law?
 At these $\beta$'s, $P(Q)$ is approximately given by gaussian form 
\begin{equation}
P(Q)\propto \exp(-\kappa_V(\beta)\ \  Q^2),
\end{equation} 
 and the fourier transform $Z(\theta)$ of this gaussian form is also 
gaussian form $Z(\theta)=\exp (-\alpha_V \theta^2)$.
 Then free energy $F(\theta)=-\frac1V \ln Z(\theta)$ is given by quadratic 
function $\theta^2$. For $L=15$ to $L=25,$ calculated $F(\theta)$ is almost 
$V$-independent. Since $Z(\theta)=\sum_{Q} P(Q) e^{i \theta Q}$ suffers 
from the statistical error $\delta P$, flattening is expected at the 
$\theta$ where the free energy becomes the order of magnitude,
\begin{equation}
      VF(\theta_f) \sim -\ln (\delta P)
\end {equation}
and we obtain the relation 
\begin{equation}
\theta_f \sim V^{-1/2}.
\end {equation}
Of course, $\delta P$ may lead to negative value for $Z(\theta)$ at 
$\theta \gtsim \theta_f$. In  $\beta=3.45$, however, $L=15, 18,$ and $ 20$ 
simulations with $N=5\times 10^5$ iterations all give positive $Z(\theta)$ 
for all $\theta( 0 \leq \theta \leq \pi)$.
 Only $L=25$ case leads to negative $Z(\theta)$ at $\theta \gtsim 0.7\pi$. 
We would expect both positive excess fluctuation and negative excess one. 
So why positive excess results were more often observed in $\beta=3.45$ 
simulations remains as a question.
 Numerical results for $Z(\theta)$ and  $F(\theta)$ is shown in the table \ref{table:F&Z}
for $\beta=3.45 $ and $L=20$ case. The value of $Z(\theta)$ in flat region 
is the same order as $\delta P \sim 1/\sqrt N \sim 1.4\times 10^{-3}( 
N=5\times 10^5).$\par
\begin{table}[h]
\caption[fr ]{The  free energy and partition function for $\beta$=3.45 and 
$L$=20.}
\label{table:F&Z}
\begin{center}
\begin{tabular}{c|c|c|c}
\hline
\hline
$\theta$  & $F(\theta)$    & $Z(\theta)$ & $\delta Z(\theta)$ \\
\hline 
 $0     $&     $    0         $   & $   1.00       $&   $1.37*10^{-3} $ \\
 $0.1\pi$&     $ 5.65*10^{-4} $   & $ 7.98*10^{-1} $&   $1.23*10^{-3} $ \\
 $0.2\pi$&     $ 2.23*10^{-3} $   & $  4.10*10^{-1}$&   $ 9.97*10^{-4}$  \\
 $0.3\pi$&     $ 4.92*10^{-3} $   & $  1.40*10^{-1}$&   $ 9.06*10^{-4}$  \\
 $0.4\pi$&     $ 8.48*10^{-3} $   & $  3.37*10^{-2}$&   $ 8.94*10^{-4}$  \\
 $0.5\pi$&     $ 1.26*10^{-2} $   & $  6.59*10^{-3}$&   $ 8.94*10^{-4}$  \\
 $0.6\pi$&     $ 1.56*10^{-2} $   & $  1.92*10^{-3}$&   $ 8.94*10^{-4}$  \\
 $0.7\pi$&     $ 1.61*10^{-2} $   & $  1.59*10^{-3}$&   $ 9.06*10^{-4}$  \\
 $0.8\pi$&     $ 1.59*10^{-2} $   & $  1.71*10^{-3}$&   $ 9.97*10^{-4}$  \\
 $0.9\pi$&     $ 1.59*10^{-2} $   & $  1.76*10^{-3}$&   $ 1.23*10^{-3}$  \\
 $\pi$&        $ 1.59*10^{-2} $   & $  1.76*10^{-3}$&   $ 1.37*10^{-3}$  \\
\hline 
\end{tabular} 
\end{center}
\end{table} 
%
\begin{table}[h]
\caption[chi]{The result of chi-square-fitting to $\ln P(Q)$ in term of
 the series starting from  $ a_0^\gamma+\sum_{n=1} a_n^ \gamma  
|Q|^{\gamma+n-1} $ for various $\beta$. }
\label{table:gamma}
\begin{center}
\begin{tabular}{l|c|c|c|c|c|c|c|c}
\hline
\hline
$\beta$ & $\chi^2$ & $\chi^2/$dof &$ \gamma $&$ a_0^\gamma $&$ 
a_1^\gamma$&$ a_2^\gamma $& $a_3^\gamma*10^{6}$ &$ a_4^\gamma*10^8$ \\
\hline
 $0$  & $24.14$ & $1.207$ &$2.08$& $-2.512$ & $-0.0194$ & 
$0.000293$ & $-9.32$ & $11.8$ \\ 
 $1$  & $22.45$ & $1.123$ &$1.96$& $-2.418$ & $-0.0260$ & 
$-0.000155$ & $3.73$ & $-3.59$ \\ 
 $2$  & $29.70$ & $1.485$ &$2.06$& $-2.267$ & $-0.0324$ &
$0.000383$ & $-11.2$ &$15.2$ \\ 
$3.45$  & $23.19$ & $1. 160$ &$2.0$& $-1.661$ & $-0. 122$ & 
$0.00261$ & $-39.1$ &  $1.84$ \\
 $4$  & $38.98$ & $1.949$ &$1.59$& $-0.927$ & $-0.607$ & $0.0104$ & 
$-251$ & $287.5$ \\ 
$5$  & $319$ & $15.950$ &$1.12$& $-0.0274$ & $-4.317$ & $-0. 0157$ & 
$-588$ &  $684$ \\ 
 $6$  & $2813$ & $140.65$ &$1.09$& $-0.00156$ & $-7.534$ & 
$-0.0209$ & $268$ &   $-1210$ \\ 
\hline
\end{tabular} 
\end{center}
\end{table}
  Once we found the flattening phenomena in rather large $\beta$ regions, we 
investigated 
 the strong coupling region again to see  whether it is the phenomenon 
peculiar to 
 weaker coupling regions. Surprisingly, similar phenomenon is found also in 
strong coupling
  regions. We show the case of  $\beta = 1.0$, which is  sufficiently 
strong; chi-square fitting 
  with gaussian form is quite good(Table.\ref{table:chi2}).  Fig.1 shown in section 1 is
 the result obtained for strong coupling region, $\beta=1.0$ and $L=10$ for 500k iterations.
Again the position of flattening is consistent with the value estimated 
from $\delta P$.
As is seen above, the calculation of the partition function with the 
(\ref{eq:pq}) using 
the observed value of $P(Q)$ directly will be called ``direct method". In 
this 
method $Z(\theta)$ receives statistical error of the order $\delta P$, and 
flattening will 
occur at $\theta_f$.
 On the other hand in the set method, the topological charge distribution 
itself  is obtained 
 over quite wide range of $Q$. We first fit  it by appropriate analytical 
 function and 
 using the fitted function $P_S (Q)$( ``smooth"), we can evaluate 
$Z(\theta)$. 
 This latter method will be called ``fitting method". 
 In strong coupling regions, the indirect method will lead to gaussian 
function 
 for $P(Q)$ and  the fourier series will again gives gaussian function of 
$\theta$ 
 for $Z(\theta)$. It gives smooth $\theta^2$ type free energy distribution 
up to 
 $\theta=\pi$. Namely no flattening phenomena will appear.\par
 Actually in $U(1)$ LGT, it is known $P(Q)$ is gaussian for all $\beta$, 
and 
 $\theta^2$ type free energy distribution up to $\theta=\pi$. But if we 
use the 
 observed $P(Q)$ directly to obtain partition function, it  will cause
  flattening phenomena. \par
%
\subsection{ Strong and weak coupling region}
As is seen in Table.\ref{table:chi2}, $\ln P(Q)$ can be fitted by gaussian 
form well up to $\beta=1.5$, 
but beyond that value the fitting becomes worse, and higher power terms are 
necessary. Up to $\beta=3.5$ quartic function fits rather well, still the 
most dominant power is 2 in this region( see 
Table.\ref{table:chi4}). 
 In $\beta \ge 4.0$ the power series up to the power 4 is
not enough as is seen from Table.\ref{table:chi4}. The value of chi-square 
become quite bad 
in $\beta \ge 4.0$. In weaker coupling region ($\beta \gtsim 4.0$), we 
tried another type of
 function, which is series starting from some power $\gamma $. 
\begin{equation}
\ln P(Q)= a_0^\gamma+\sum_{n=1} a_n^ \gamma  |Q|^{\gamma+n-1} 
\end{equation}
 This drastically improves the 
 chi-square fit in $4.0 \gtsim \beta \gtsim 5.0$ (see 
Table.\ref{table:gamma}. and compare with Table.\ref{table:chi4}.). 
 The chi-square value for $\beta=4$ was 3497 in Table.\ref{table:chi4},
  it becomes 38.98 in Table.\ref{table:gamma}.
 From Table.\ref{table:gamma}, 
 we understand that leading power $\gamma$ is 2 for $\beta \ltsim 3.5$ and the 
leading 
 power becomes 1 in $\beta \gtsim 5$, thus we will call this region 
weak coupling region.
  $\beta =4.0$ is ``intermediate", since leading power $\gamma $ is 
 about 1.59, which is just medium value of $\gamma =$2(gaussian in strong  coupling regions) and $\gamma $=1(weak coupling regions). \par
 In weak coupling regions excitation of topological charge is suppressed 
stronger as $\beta$ becomes larger. For example, 
\begin{equation}
\frac{P(2)}{P(1)}= \frac{1}{1.12},\quad \frac{1}{3.3},\quad \frac{1}{126}
\end{equation}
for $\beta=2.0, 4.0, 5.0$ respectively. \par
On the contrary large topological charge configuration is relatively 
important and 
gaussian distribution is obtained in strong coupling region. It leads to 
the partition function, 
$\vartheta_3(\nu, \tau )$, third elliptic theta function. Here $\nu = 
\theta/({2 \pi}), \tau = i \kappa_V(\beta)/\pi $.
 Note that $\theta $ is extended to complex 
values. 
This function clearly shows 
the distribution of partition function zeros leading to the first order 
phase 
transition at $\theta=\pi$.\par
  In weak coupling regions large topological 
charge 
contribution is suppressed. In this region $-\ln P(Q)$ is given 
approximately 
by $a^{\gamma}_1 |Q|$( with $\gamma \sim 1.0$). This form of $P(Q)$ leads 
to the partition function given by 
the generating function of Chebyshev Polynomials.
\begin{eqnarray}
Z(\theta)&=& \sum ^{\infty}_{Q=-\infty } c^{|Q|} e^{i\theta Q} \\
         &=& \frac{1- c^2}{1-2 c \cos \theta +c^2}
\end{eqnarray}
In weak coupling regions, the value of $c$ is quite small and $c^2$ in 
this equation can be
discarded. It leads to the expectation value of topological charge almost 
given by $\sin \theta$ form.
\begin{eqnarray}
i<Q> =  \frac1V \frac{dZ}{d \theta}/Z \sim \frac{2 c}{V}   \sin \theta 
 \end{eqnarray}
when $|c|<<1$. At $\beta=6.0, L=20$, the value of $c$ is $c=1.34\times 
10^{-4}, 2c/V \sim 6.68\times 10^{-7}$.
Actually calculated $ i <Q>$ shows sine function type behavior over whole 
range of $\theta$ in weak coupling regions. No discrete phase transition 
is expected at $\theta=\pi$ in weak coupling regions.
Whether there is a continuous phase transition( second order one) or not 
needs further
precise investigation such as study of correlation length.  \par

\section{Conclusoins and discussion}

In this paper we have investigated two dimensional $CP^2$ model with 
$\theta$-term numerically. Topological charge distribution is measured by 
Monte Carlo method based on set method and trial function method. Then it 
is used to obtain partition function $Z(\theta)$ as a function of theta 
parameter. \par

Topological charge distribution $P(Q)$ is gaussian in strong coupling 
regions( $\beta \ltsim 1.5$). In $\beta =2 \sim 3.5$ region, the exponent 
of $P(Q)$ can not be simply given by $Q^2$ but additional higher powers 
are necessary(Table I and II). Still the leading contribution comes from 
 $Q^2$ term(Table V).\par
In weak coupling regions,  i.e., $\beta \gtsim 4.5$, the leading power of 
exponent of $P(Q)$ is $|Q|$. Sharp decrease of $P(Q)$ at larger $|Q|$ is 
observed.\par
  At $\beta =4$, which is intermediate of strong and weak 
coupling region is characterized by leading power of exponent of  $P(Q)$ 
is given by $|Q|^{1.59}$, whose exponent is the intermediate value of $Q^2$( gaussian 
$P(Q)$ ) in strong coupling region and $|Q|^1$ in weak coupling region. 
\par
Free energy as a function of $\theta$ is investigated. We observed 
``flattening " of $\theta$-distribution of free energy at $\theta_f$. In 
our analysis, the main reason of this flattening is attributed to the 
statistical fluctuation $\delta P$ of $P(Q)$ at $Q=0$ and   $\delta P \sim 
O(1/\sqrt N )$, where $N$ is the number of measurements. Volume dependence 
at each $\beta$ is explained by this interpretation.  \par
  Topological charge distribution $P(Q)$ is measured over wide range of $Q$. 
But direct method to obtain $Z(\theta)$ through fourier series with direct 
use of measured topological charge distribution $P(Q)$ leads to flattening 
caused by $\delta P$. We tried to obtain approximate form of $P(Q)$ by 
fitting method. The smooth function $P(Q)_S$ obtained by fitting and used 
to evaluate $Z(\theta)_S$ has following properties.    
In strong coupling region,  the gaussian topological charge distribution 
leads to $Z(\theta)_S$ with the third elliptic theta function 
$\vartheta _3$ and it leads to infinite number of zeros  of partition 
function leading to the first order phase transition at $\theta=\pi$ in 
infinite volume limit. 
In weak coupling region, $Z(\theta)_S$ is approximated by the functional form

\begin{eqnarray}
Z(\theta)_S =  \frac{1-c^2}{1-2 \cos \theta +c^2}, \quad (c<<1)
 \end{eqnarray}
which is the generating function of Chebyshev polynomials. Due to this 
form, the expectation value of topological charge
\begin{eqnarray}
i <Q>\propto \sin \theta
 \end{eqnarray}
 No first order phase transition at $\theta=\pi $ is  expected. 

\section*{Acknowledgements}
We would like to thank R. Burkhalter for stimulating discussions.
\appendix
\section{Integration Measure}
%
In this appendix, we will discuss briefly about the measure of integration.
The measure is defined as
\begin{eqnarray}
dz_1 dz_2 dz_3 \delta(|z_1|^2 + |z_2|^2+|z_3|^2 -1).
\end{eqnarray}
The complex numbers $z_i(i=1,2,3)$ are given by two real numbers $x_i ,y_i$
as $  z_i = x_i + i y_i   $.
The measure becomes
\begin{eqnarray}
dx_1 dy_1 dx_2 dy_2 dz_3 dz_3 \delta(x_1^2 + y_2^2+ x_2^2 +y_2^2
                                  + x_3^2 + y_3^2 -1).
\end{eqnarray}
When the parameters   $ x_i ,y_i (i = 1,2,3)$ are changed
                     to polar coordinates $(k_i,a_i(i = 1,2,3))$
\begin{eqnarray}
x_i &=& k_i \cos a_i, \\
y_i &=& k_i \sin a_i,
\end{eqnarray}
where $  k_i \ge 0$ and $ 0\le a_i \le 2 \pi$.
The measure becomes
\begin{eqnarray}
k_1 \  k_2 \  k_3\  dk_1\  dk_2\  dk_3 \ \delta(k_1^2 +k_2^2 + k_3^2 -1) \ da_1\  da_2 \ da_3 .
\end{eqnarray}
We translate the parameters 
from $ k_1, k_2, k_3 $ to 3 dimensional polar coordinates $(l,t_1,t_2)$
\begin{eqnarray}
k_1 &=& l \sin t_1 \cos t_2 ,  \\
k_2 &=& l \sin t_1 \sin t_2,\\ 
k_3 &=& l \cos t_2 ,
\end{eqnarray}
\begin{eqnarray}
l^5 \ \delta(l^2-1) \sin^3 t_1 \cos t_1 dt_1 \sin t_2 \cos t_2 \ dl \ da_1 \  da_2 \ da_3,
\end{eqnarray}
where $ l \ge 0$ and $ 0\le t_1, t_2 \le \frac{\pi}{2}$. 
So the measure which we use in Monte Carlo simulation becomes finally
\begin{eqnarray}
dm \ dn \ da_1 \ da_2 \ da_3,
\end{eqnarray}
after integration about $l$ is done. The variables  $m, n$ are given by
\begin{eqnarray}
m &=& \frac{1}{8} (\cos 2 t_2 - \frac{1}{4} \cos 4 t_1), \\
n &=& \frac{1}{4} \cos 2 t_2.
\end{eqnarray}
\vfill
\eject

\end{document}